\documentclass{osa-article}

\journal{oe}

\usepackage{lineno}
 \usepackage[markup=underlined]{changes}
\definechangesauthor[color=red, name={Adam Linek}]{AL} 
\definechangesauthor[color=blue, name={Marcin Witkowski}]{mw} 

\usepackage{todonotes}
\makeatletter
\setcommentmarkup{%
  \todo[
    textcolor={authorcolor!70!black},
    backgroundcolor={authorcolor!20!white},
    linecolor={authorcolor!30!white},
    size=\scriptsize]%
  {\@nameuse{Changes@AuthorName#2}\\#1}
}
\makeatother

\newcommand{\printtransition}{$^1$S$_0$\makebox[15pt]{$\to$}$^3$P$_1~$}
\begin{document}

\title{Absolute frequency measurement of the $6s^2~^1S_0 \to 6s6p~^3P_1$ $F=3/2\to F'=5/2$ $^{\text{201}}$Hg transition with background-free saturation spectroscopy}

\author{Adam Linek\authormark{*}, Piotr Morzyński, and Marcin Witkowski}

\address{Institute of Physics, Faculty of Physics, Astronomy and Informatics, Nicolaus Copernicus University, Grudzi\c{a}dzka 5, PL-87-100 Toru\'n, Poland}

\email{\authormark{*}a.linek@doktorant.umk.pl} 



\begin{abstract}
We report the development of a method for eliminating background-induced systematic shifts affecting precise measurements of saturation absorption signals. With this technique,
we measured the absolute frequency of the $6s^2~^1\text{S}_0 \to 6s6p~^3\text{P}_1$ transition in $^{201a}\text{Hg}$ ($F=3/2\to F'=5/2$) to be \textcolor{black}{$1181541111051(83)$~kHz}. The measurement was referenced  with an optical frequency comb synchronized to the frequency of the local representation of the UTC.
This specific atomic line is situated on the steep slope of the Doppler background at room temperature, which results in 
a
frequency systematic shift.  We determined the dependence of this shift on the properties of both the spectral line and the background of the measured signal.  
\end{abstract}

\section{Introduction}
Saturation spectroscopy is a powerful Doppler-free technique providing detailed information about the electronic structure of atoms through very precise frequency measurements of the atomic transitions. The method is the basis for various laser frequency stabilization and locking techniques~\cite{Raj,Bjorklund,Debs,Ooijen,Gawlik,Krzemien,Aldous}. 
    The saturation spectrum's
high-frequency resolution 
is achieved thanks to 
a
significant reduction of the Doppler broadening.  
However, the spectral lines obtained with the Doppler-free saturation spectroscopy are often situated on a non-flat Doppler background, which limits the precision of the measurements. \textcolor{black}{This limitation is because 
the signal's maximum no longer corresponds to the atomic transition frequency.}
\textcolor{black}{This frequency shift is usually estimated from the fit to the measured line and background profiles. Such treatment is sufficient enough for high-quality data and fitting models.
However, the low signal-to-noise ratio measurements result in high uncertainties of the fitting parameters, which makes this method inefficient.
}

\begin{figure}[ht]
    \centering
    \includegraphics[width=1\linewidth]{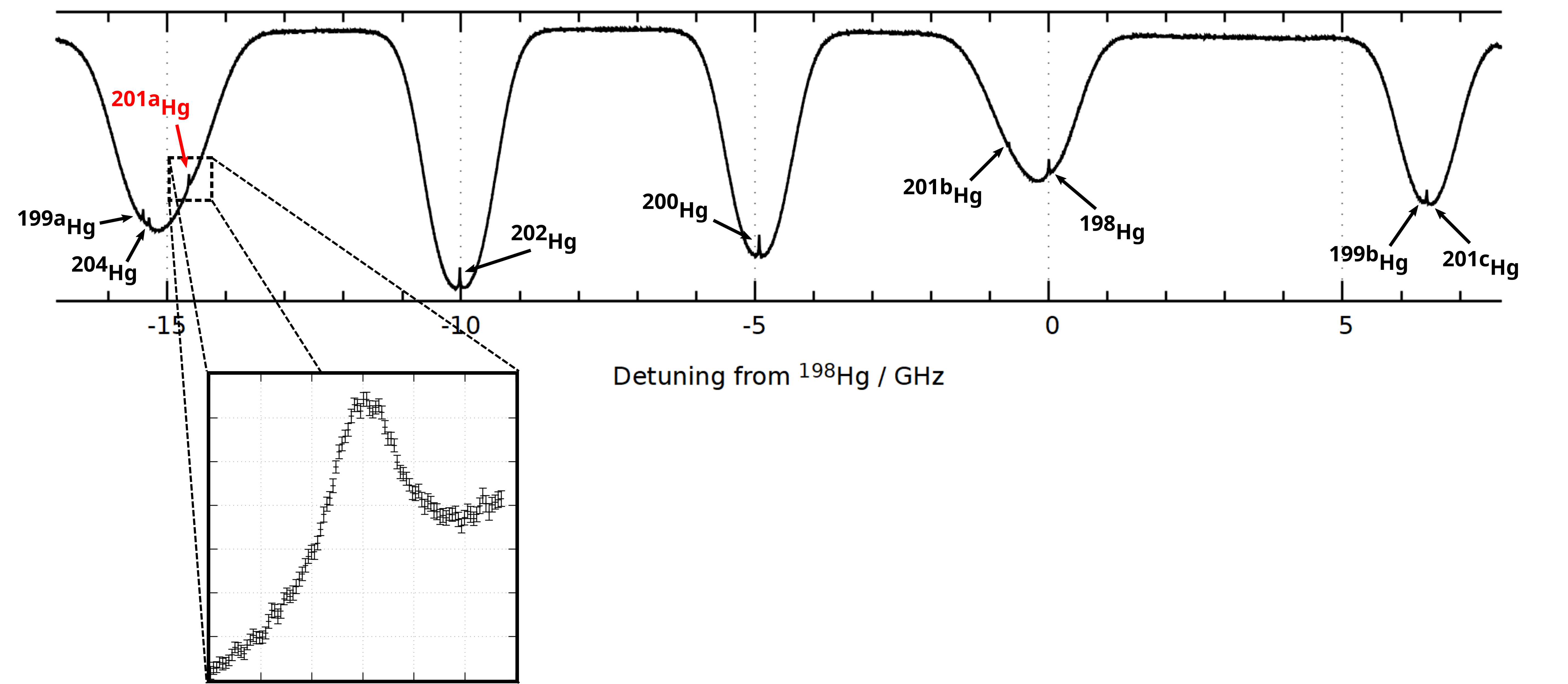}
    \caption{\textcolor{black}{The saturation spectroscopy of the \printtransition transition 
    in the room-temperature mercury vapor cell. Five 
    Doppler-broadened profiles of the mercury isotopes with sub-Doppler Lamb dips are visible. 
    At room temperature, 
    the Doppler-broadened profiles are of $\sim 1$~GHz widths, much more than the
    isotopic shifts of some of the Lamb dips. 
    The overlapping leads to situating the sub-Doppler spectral lines at the slopes, while the one for the $^{201a}$Hg (shown in the inset) is the steepest. 
    }}
    \label{fig:widmo}
\end{figure}

In this paper, we present a novel method stemming from saturation spectroscopy but modified by employing a new way to deal with the undesirable background-induced systematic shift of the atomic transition frequency. The technique is fairly insensitive to temperature changes, which entail changes in the Doppler profile and, thus, in the shape of the measured signal's background. Using this method, we measured the absolute frequency of the $6s^2~^1\text{S}_0 \to 6s6p~^3\text{P}_1$ transition in $^{201a}\text{Hg}$ ($F=3/2\to F'=5/2$) with a room-temperature mercury vapor cell. This line lies on the steep slope of the Doppler profile at room temperature (Fig.~\ref{fig:widmo}), which made it possible to exploit the method's advantages fully. The uncertainty budget of the measurement includes all significant frequency shifting effects such as AC-Stark, Zeeman shift, and pressure-dependent collisional shift.

The study of spectral lines of mercury has been a subject of interest for over a century~\cite{Runge1891,Moore1958,Stager1960,Schweitzer1963,Reader1980,Kohler1961,Saloman2006,Sansonetti2010,Kramida2011}. \textcolor{black}{Although the mercury's intercombination \printtransition line has been the subject of numerous spectroscopic studies (this applies mainly to isotope $^{198}$Hg), the 
$F=3/2\to F'=5/2$ $^{\text{201}}$Hg transition, to our knowledge, has not been measured directly. There were, however, reported results~\cite{Schweitzer1963}
of its isotope shift relative to $^{198}$Hg (measured in~\cite{Schweitzer1963a}), from which one can infer the frequency value to be $1181541128(17)$~MHz. This value agrees with our result $1181541111051(83)$~kHz within the uncertainty.}
The advent of high-power diode lasers and a non-linear frequency conversion paved the way for more precise spectroscopic measurements of mercury~\cite{Witkowski1} and enabled the exploration of new areas of fundamental research based on the mercury atom~\cite{Safronova2018,Bouchiat1977,Latha2009,Swallows2013,Graner2016,Angstmann2004}.
Recently, the isotopically resolved Hg spectrum of the $^1S_0 \to ~^3P_1$ intercombination transition was used for developing a new method for measuring the gaseous elemental mercury concentration in the air ~\cite{Hodges}. Our measurement enriches this spectrum with a new value, which will \textcolor{black}{increase} the method's accuracy. 


\section{Experimental setup}
\begin{figure}[ht]
    \centering
    \includegraphics[width=1\linewidth]{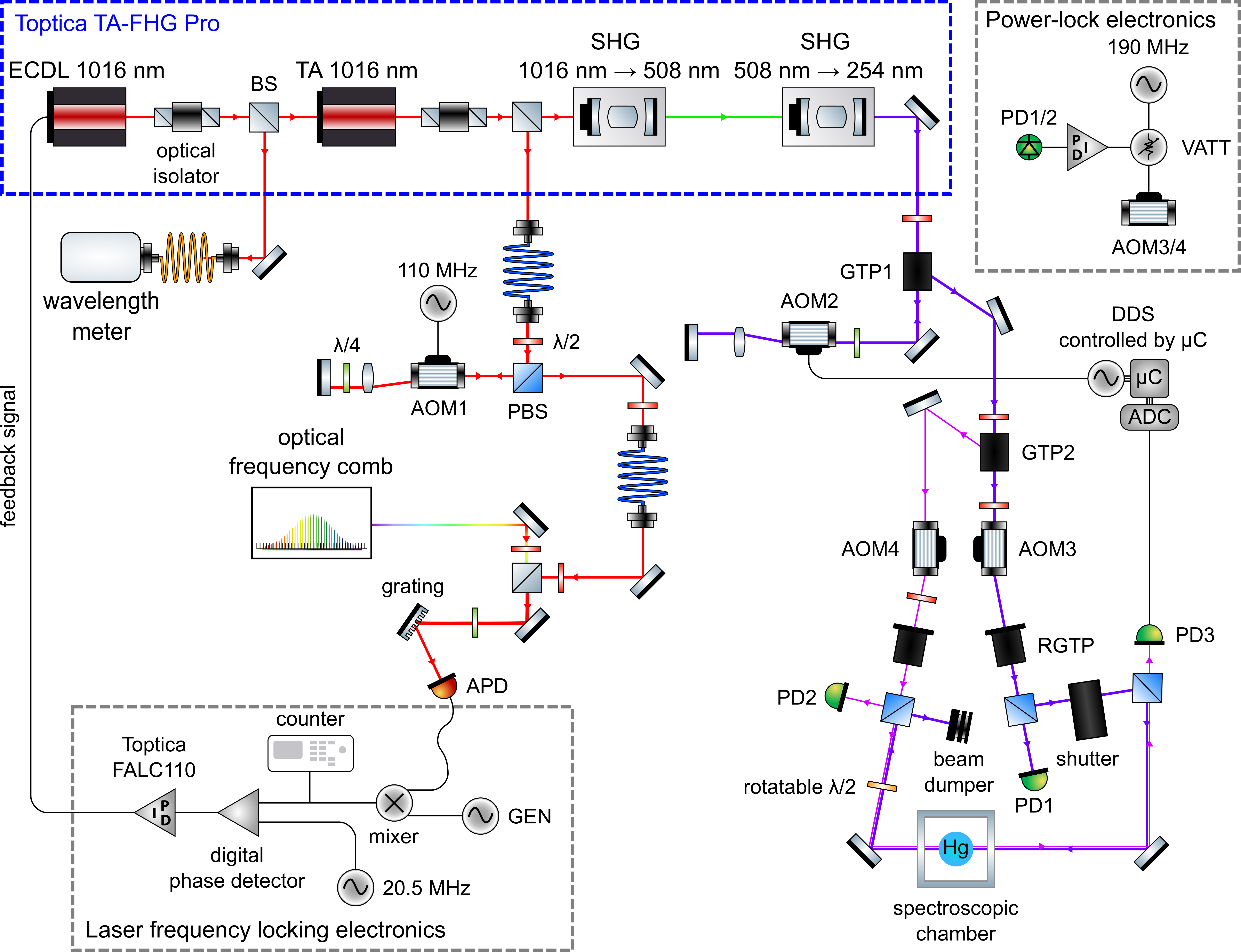}
    \caption{Scheme of the experimental setup consisting of a     laser source (Toptica TA-FHG Pro), an optical frequency comb, and a saturation spectroscopy setup. The infrared 1016~nm laser beam generated by an ECDL is referenced to the optical frequency comb. The frequency of IR light is doubled twice to obtain the UV light at a wavelength of $254$~nm. The spectroscopy setup consists of a spectroscopic mercury cell, an AOM2 in double-pass configuration, and two independent power stabilization systems for the pumping and probing beams.}
    \label{fig:setup}
\end{figure}

The experimental setup scheme is shown in Fig.~\ref{fig:setup}. The laser system (Toptica TA-FHG Pro) is based on a $1016$~nm laser diode in an external resonator configuration (ECDL). The power of the infrared light is amplified up to $2$~W with the tapered amplifier (TA). The frequency is doubled twice, resulting in the ultraviolet 254~nm laser beam with a power of up to $120$~mW. \textcolor{black}{The typical laser beam intensity used in the experiment is $140$~mW/cm$^2$}.

To stabilize and measure the frequency of the ECDL, an optical frequency comb is used. After amplification and frequency shifting by an acousto-optic modulator (AOM1), a part of the infrared light is spatially superimposed with the comb output beam and sent to an avalanche photodiode (APD) for the frequency beat detection. The beat note signal is mixed with the RF signal from the generator (GEN) so that the frequency of the resulting signal is 
down-converted
and adapted to the operating range of the RF counter. Its readings are used to determine the absolute frequency of infrared, and thus ultraviolet, laser light. In addition, the mixed RF signal is provided to a digital phase detector, which outputs a DC signal proportional to the phase difference between the mixed RF signal and a reference $20.5$~MHz signal. The DC output signal is provided to the PID controller (Toptica FALC110), which generates a signal modulating the ECDL's current and its piezo voltage. With this approach, the infrared laser light is frequency-narrowed to about $100$~kHz and locked to the optical frequency comb mode. To determine the comb mode number, a small fraction of the infrared light is guided to the wavelength meter, whose uncertainty ($60$~MHz) is a few times smaller than the comb repetition frequency ($250$~MHz).

Ultraviolet light is sent to the spectroscopy system. Using the saturation effect of the atomic transition~\cite{Preston}, a \textcolor{black}{sub-Doppler Lamb dip} 
is
observed. \textcolor{black}{While the spectral line is free from the Doppler broadening, its width is a few times larger than the natural linewidth due to 
power broadening, transient-time broadening, and pressure broadening.} The spectroscopy signal is used as an input of the frequency stabilization system, which tunes the UV frequency to the center of the atomic line. The frequency shifting is realized with an acousto-optic modulator (AOM2), whose RF driving signal is generated by a \textmu C-operated direct digital synthesizer (DDS). The UV frequency-shifted beam is separated with a Glan-Taylor prism (GTP1) and split into 
another two beams by using the
prism (GTP2). These two beams act as the pumping and probing beams in the spectroscopy system. Their power ratio can be chosen with a half-wave plate in front of the GTP2.
We applied both beams' independent power-lock systems with PD1 and PD2
to ensure good power stability. \textcolor{black}{For typical experimental conditions, the relative power stability is at the level of $8.3$~ppm and $6.5$~ppm for the pumping and probing beam, respectively.}
To eliminate the influence of the Doppler background, we periodically block the pumping beam, which disables the saturation of the atomic transition and measure the background level, which is then digitally subtracted. Pumping and probing beams are sent through AR-coated wedged fused silica windows to a 1 mm thick cylindrical absorption cell containing Hg vapors. The cell is attached to a servo that allows its rotation, thus changing the length of the optical path of the probe beam inside the cell. The saturated absorption spectroscopy signal is detected with a photodiode (PD3) and sent to the microcontroller, where the UV frequency correction is calculated (see \textcolor{black}{section}~\ref{sec:metoda}).

\begin{figure}[ht]
    \centering
    \includegraphics[width=0.7\linewidth]{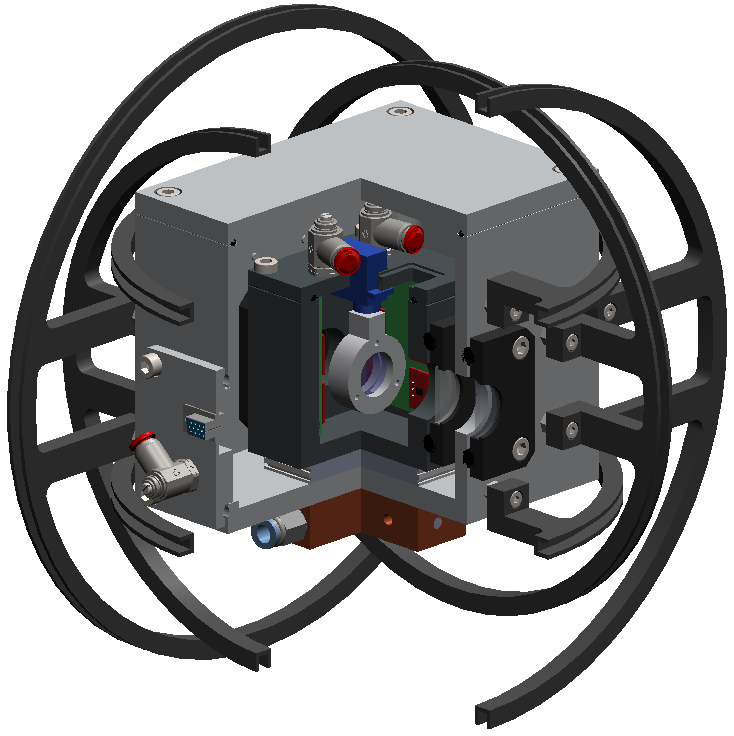}
    \caption{Quarter-section of the spectroscopic chamber design. Two aluminum chambers surround the spectroscopic. The temperature stabilization is realized with two Peltier modules below the inner chamber. The excessive heat is removed from the system by coolant flowing through a copper plate. The cell is surrounded by eight offset-compensated 3-axis magnetometers enabling magnetic field and magnetic gradient measurements. 
The chamber is equipped with six coils in three independent Helmholtz arrangements 
to compensate for a stray magnetic field.
UV light passes through the chamber  via four AR-coated wedged fused silica windows.}
    \label{fig:chamber}
\end{figure}

To account for the influence of the Zeeman effect and collision-induced shifts, the spectroscopic cell was placed inside a spectroscopic chamber shown in Fig.~\ref{fig:chamber}.
The chamber enables thermal stabilization at the level of $0.1$\textcolor{black}{$~^\circ$C}.
\textcolor{black}{The temperature reading is based on a measurement of a Pt100's resistance with the use of an analog-digital converter. The model used for translating the resistance into the temperature extends its uncertainty to $0.5~^\circ$C.}
The chamber is equipped with three pairs of coils in Helmholtz configuration to compensate for a stray magnetic field.
Eight 3-axis offset-compensated magnetometers distributed around the cell enable both magnetic field strength (with the uncertainty of a few \textmu T) and magnetic gradient measurements.

The whole 
spectroscopic chamber consists of two aluminum chambers: the inner and the outer one depicted in Fig.~\ref{fig:chamber} in grey and silver, respectively. The volume between the inner and outer chambers is pumped out to improve thermal insulation. The inner chamber is filled with nitrogen, 
preventing
unwanted water condensation or freezing during chamber cooling. 

The temperature stabilization is realized with two Peltier modules below the inner chamber. The excessive heat is removed through the water-cooled copper plate. The temperature of the inner chamber is measured using a Pt100 temperature sensor attached to its wall. This provides 
feedback to the PI controller.

\section{Method}
\label{sec:metoda}
\begin{figure}[ht]
    \centering
    \includegraphics[width=0.65\linewidth]{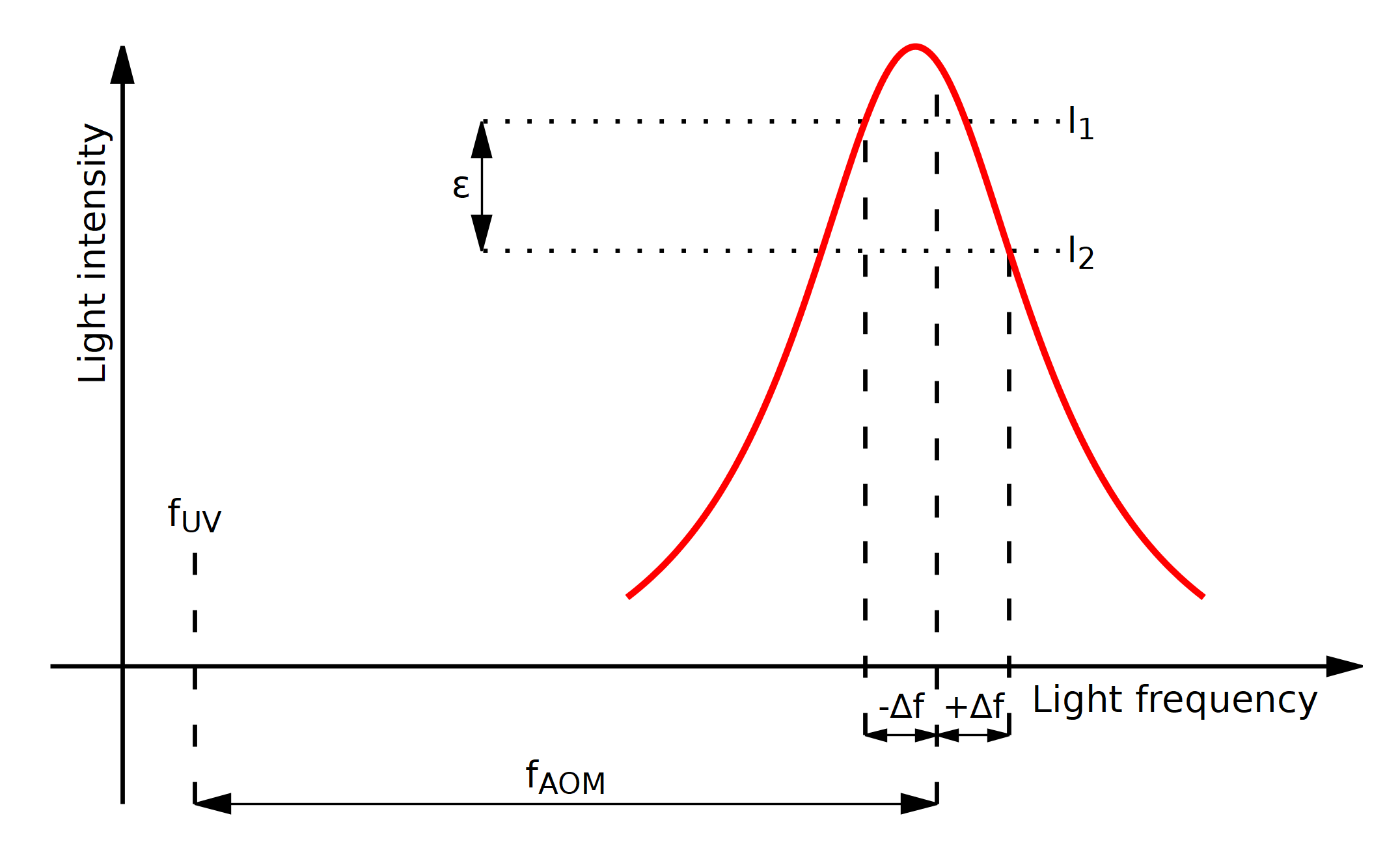}
    \caption{Scheme of the DLM. The frequency of the laser light is shifted cyclically by $\Delta f$ towards lower and higher values corresponding to the opposite slopes of the \textcolor{black}{sub-Doppler} spectral line \textcolor{black}{obtained with the pumping beam enabled}. After each frequency jump, the spectroscopic signal is measured, and the difference between two consecutive values $\varepsilon = I_1 - I_2$ is used to calculate the frequency correction by which the $f_\text{AOM}$ is shifted after each cycle.}
    \label{fig:lock1}
\end{figure}

 To account for the background-induced frequency shift, we improved a digital lock method (DLM), 
 which has
 proven to work very well for symmetric spectral lines~\cite{Morzynski1,Morzynski2,Witkowski1}. The scheme of this method is shown in Fig.~\ref{fig:lock1}. 
 The DLM is based on cyclical switching 
 of
 the laser frequency between two values $f_{\text{UV}}+f_{\text{AOM}}-\Delta f$ and $f_{\text{UV}}+f_{\text{AOM}}+\Delta f$, which are related to opposite slopes of a measured line. After each frequency jump, the spectroscopy signal $I$ is measured by a photodiode. The difference between two consecutive values $(\varepsilon=I_1-I_2=I_{f_{\text{UV}}+f_{\text{AOM}} - \Delta f} - I_{f_{\text{UV}}+f_{\text{AOM}} + \Delta f})$ is used to calculate the frequency correction $f_{corr}$
\begin{equation}
    f_{corr} = \left(I_{f_\text{UV}+f_{\text{AOM}} - \Delta f} - I_{f_{\text{UV}}+f_{\text{AOM}} + \Delta f}\right)\cdot \text{gain},
\end{equation}
by which the $f_{\text{AOM}}$ is shifted after each cycle. The \textmu C calculates the frequency correction after every two frequency jumps. For the symmetric line, the laser frequency \textcolor{black}{lying} exactly at the center of the measured profile results in $\varepsilon=0$, thus also $f_{corr}=0$. With this loop, the laser frequency $\left(f_\text{UV}+f_\text{AOM}\right)$ follows the spectral line's maximum position.

\begin{figure}[ht]
    \centering
    \includegraphics[width=0.65\linewidth]{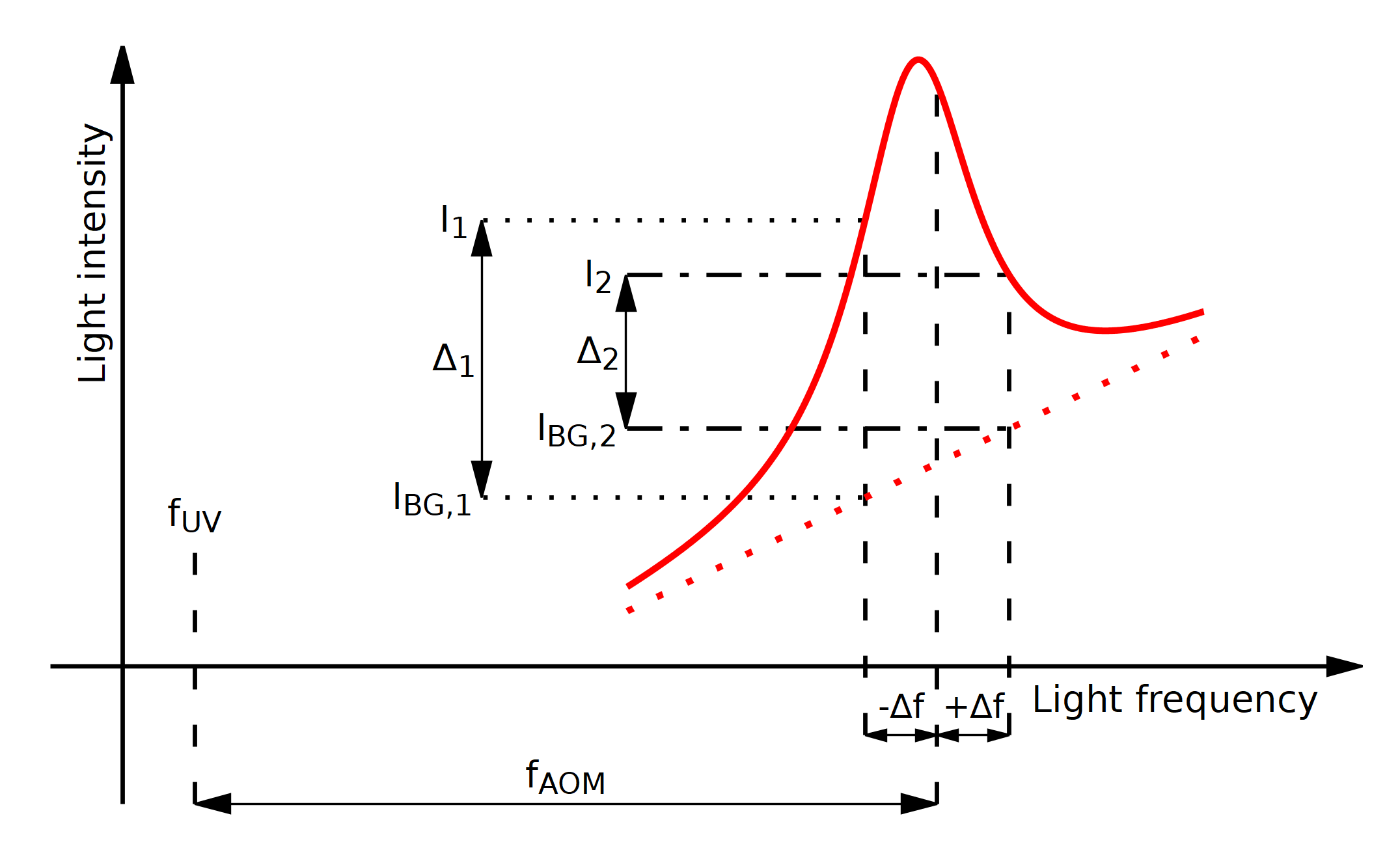}
    \caption{Scheme of the mDLM.
     \textcolor{black}{At the beginning of the cycle, the pumping beam is disabled, and the background-only measurement (dashed red line) is performed at $f_\text{UV}+f_\text{AOM} \pm \Delta f$, giving $I_{BG, 1}, I_{BG, 2}$. Then, the pumping beam is switched on and the spectroscopic signal of the sub-Doppler spectral line (solid red line) is measured for the same frequencies, giving $I_{1}$ and $I_{2}$.}
     The differences $\Delta_1 = I_1 - I_{BG,1}$ and $\Delta_2 = I_2 - I_{BG,2}$ are used to calculate the frequency correction by which the $f_\text{AOM}$ is shifted.}
    \label{fig:lock2}
\end{figure}

The DLM works very well for symmetrical profile. However, it may lie on a slope when a spectral line shares a Doppler-broadened absorption profile with other spectral lines, such as those from other isotopes.
The steeper the slope, the more systematic error will be introduced by the DLM (we describe this effect in detail in section~\ref{sec:systematic_shift}). To solve this issue, we developed the modified digital lock method (mDLM). The scheme of the mDLM is depicted in Fig.~\ref{fig:lock2}.
As a significant improvement, we introduced the on-the-fly background-eliminating scheme based on disabling the saturation of the atomic transition. It is managed by repeating the frequency jumping cycle with the blocked pumping beam. Once the pumping beam is disabled, the Lamb dip disappears, and only the \textcolor{black}{Doppler-broadened absorption profile} remains. The cycle of two frequency jumps and signal measurements is repeated, but this time it corresponds only to the background levels, subtracted from the previously obtained spectral line measurements. As a result, we get two values $I_{BG,1}$, $I_{BG,2}$ related to the opposite background-free slopes and use them to calculate the frequency correction $f_{corr}$
\begin{equation}
\label{eq:freq_corr}
f_{corr} = \left( \Delta_1 - \Delta_2 \right)\cdot \text{gain},
\end{equation}
where
\begin{equation*}
    \begin{split}
    \Delta_1 &= I_1 - I_{BG,1}= I^\text{line+bg}_{f_\text{UV}+f_{\text{AOM}} - \Delta f} - I^\text{bg}_{f_\text{UV}+f_{\text{AOM}} - \Delta f},\\
    \Delta_2 &= I_2 - I_{BG,2}= I^\text{line+bg}_{f_\text{UV}+f_{\text{AOM}} + \Delta f} - I^\text{bg}_{f_\text{UV}+f_{\text{AOM}} + \Delta f}.
    \end{split}
\end{equation*}
The $I^\text{line+bg}$ and $I^\text{bg}$ correspond to the values measured in the presence and absence of the pumping beam, respectively.

\begin{figure}[ht]
    \centering
    \includegraphics[width=1\linewidth]{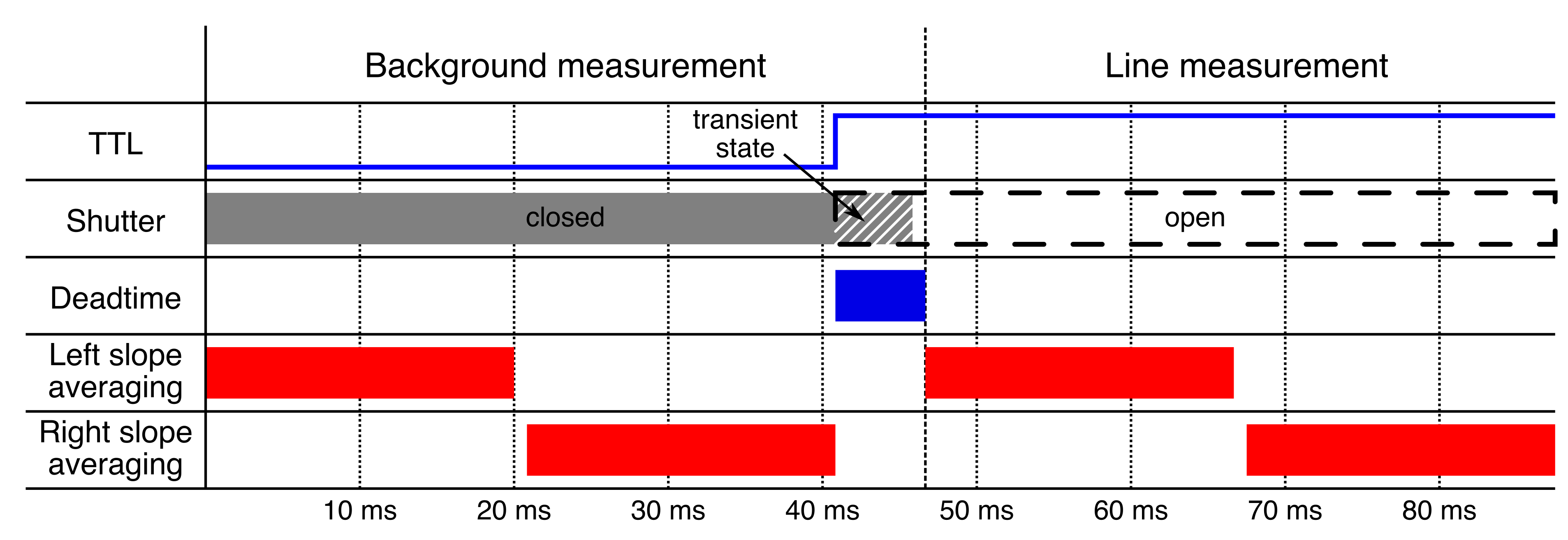}
    \caption{The time frame of the mDLM. At the beginning of the cycle, the shutter is closed, and the background level measurements are performed consecutively for the left and right slopes of the spectral line. Next, the \textmu C sends the opening signal to the shutter controller. The full-opening time is $5$~ms. Once the shutter is fully opened, the spectral line measurements are performed, and the frequency correction is calculated and applied. 
	    Afterward,
    the shutter is closed, and the cycle is repeated. 
    }
    \label{fig:timeline}
\end{figure}

Switching off the pumping beam is accomplished with a mechanical shutter controlled by a TTL signal from the \textmu C. The full-closing time of the shutter is $5$~ms. The time frame of the entire measurement cycle of the mDLM is shown in Fig.~\ref{fig:timeline}

\section{Microcontroller and direct digital synthesizer}
\begin{figure}[ht]
    \centering
    \includegraphics[width=0.7\linewidth]{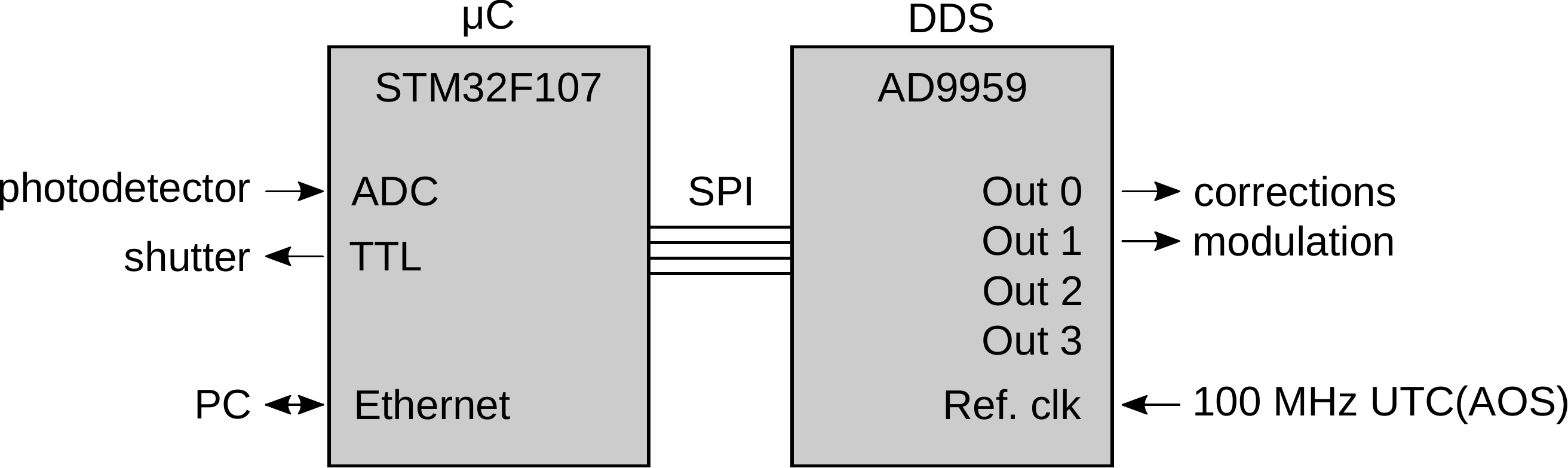}
    \caption{Scheme of the digital device for background-free spectroscopy control. The microcontroller executes the frequency correction algorithm and communicates with peripheral devices: ADC - analog-digital converter for measuring the signal from 
the
photodetector, DDS - direct digital synthesizer for precise frequency control of AOMs, digital ports for sending TTL signals, e.g.,
to the shutter. Ethernet communication acts as a user interface with the device.}
    \label{fig:DDSuC}
\end{figure}

A home-made digital device (Fig. \ref{fig:DDSuC}) consisting of a direct digital synthesizer Analog Devices AD9959 evaluation board (DDS) and a Kamami ZL26ARM evaluation board with an STM32F107 microcontroller (\textmu C) were used to precisely control the frequency of the RF signal supplied to the acousto-optical modulator and to implement the frequency correction algorithm described in section~\ref{sec:metoda}. The reference clock for DDS is a 100~MHz signal coming from UTC (AOS)~\cite{AOS, UTC}. The DDS includes four output RF channels digitally tunable from 0~MHz to 210~MHz with around 100~mHz resolution. All DDS parameters are set by the \textmu C through the SPI interface, allowing the output frequency to be switched in a microsecond which is negligible compared to the cycle time. The opening and closing of the shutter are controlled by the TTL signal from the \textmu C digital output. The signal is sent to the TTL input of the shutter driver.
  
The signal from the photodetector (PD3 in Fig.~\ref{fig:setup}) is measured using a 12-bit ADC converter built in the \textmu C. The reading from the photodiode is averaged over the multiple measurements (about a thousand points per 1~ms) to improve the signal-to-noise ratio.

The digital background-free locking algorithm is implemented in 
a
\textmu C program written in C language. The program can also be switched to the background-free or background-included scan mode to observe the line profile and determine the initial lock parameters.
All parameters of the \textmu C operation are set through commands sent via the Ethernet network. The \textmu C can also save user-selected parameters and measurements and share them via Ethernet. 

\section{Background-induced systematic shift}
\label{sec:systematic_shift}
If the separation of the spectral lines is small compared to their Doppler widths, the absorption profiles overlap, and the associated Lamb dips are no longer situated at the profile center but on the slope. 
    Consequently,
the peak of the Lamb dip is frequency-shifted according to the atomic resonant frequency. To analyze this shift, we model the measured atomic transition as a Lorentz profile with a linear background. The latter is justified since the width of \textcolor{black}{the Doppler-broadened} absorption profile is much larger than the linewidth of the atomic transition. The observed spectral line intensity profile can be then written as
\begin{equation}
    \label{eq:lorentz_z_liniowym_tlem}
    I(f) = \frac{A}{1 + \left[ \frac{2\left( f - f_0 \right)}{w} \right]^2} + a (f-f_0) + b,
\end{equation}
where $A$ and $w$ are the amplitude and the width of the Lorentz profile, respectively, $f-f_0$ is the detuning from the atomic resonant frequency, and $a$, together with $b$, is the linear slope parameters. In the case of the DLM, the value $f_\text{DLM}$ at which the light frequency is stabilized does not correspond to the resonant frequency $f_0$ of the atomic transition but is shifted instead. Moreover, this shift depends on the jumps magnitude $\Delta f$ in the frequency correction cycles. To calculate the frequency shift ($f_\text{DLM} - f_0$) as a function of $\Delta f$, one can solve the equation
\begin{equation}
    \label{eq:rownanie_na_stabilizacje}
    I(f + \Delta f) = I(f - \Delta f),
\end{equation}
which can be reduced to
\begin{equation}
    \alpha_5(f-f_0)^4 + \alpha_3(f-f_0)^2 + \alpha_2(f-f_0) + \alpha_1 = 0,
\end{equation}
where
\begin{equation*}
\begin{split}
    \alpha_5 &= 16,\\
    \alpha_3 &= 8w^2 -32\Delta f^2,\\
    \alpha_2 &= -\frac{8w^2A}{a},\\
    \alpha_1 &= w^4 + 8w^2\Delta f^2 + 16\Delta f^4.
\end{split}
\end{equation*}
The dependence of the frequency $f_\text{DLM}$ satisfying the Eq.~(\ref{eq:rownanie_na_stabilizacje}) on $\Delta f$ is elaborate but for $\Delta f < w$ can be approximated as
\begin{equation}
    \label{eq:zaleznosc_od_jrange}
    f_\text{DLM} (\Delta f) = \beta_4 (\Delta f)^4 + \beta_2 (\Delta f)^2 + \beta_1,
\end{equation}
where $\beta_i$ consist of the profile and slope parameters.

\begin{figure}[ht]
    \centering
    \includegraphics[width=0.6\linewidth]{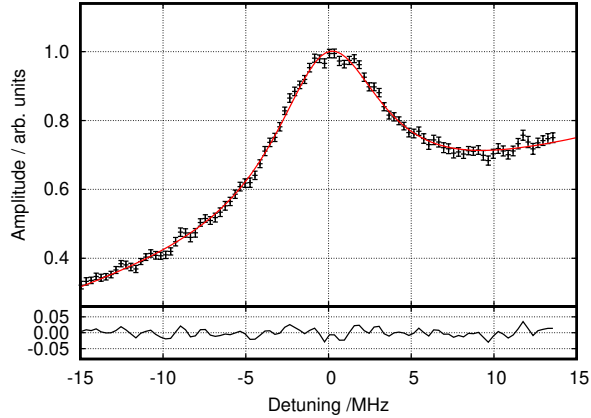}
    \caption{\textcolor{black}{Sub-Doppler spectral} line profile \textcolor{black}{obtained with the saturation spectroscopy} of the \printtransition transition in $^{201a}$Hg. This line is situated on the slope arising from sharing 
the
Doppler-broadened absorption profile with neighboring $^{199a}$Hg and $^{204}$Hg spectral lines. The results are averaged over fifteen scans and fit according to Eq.~\ref{eq:lorentz_z_liniowym_tlem}. The bottom part of the figure shows the residuals.}
    \label{fig:bg}
\end{figure}

\begin{figure}[ht]
    \centering
    \includegraphics[width=0.6\linewidth]{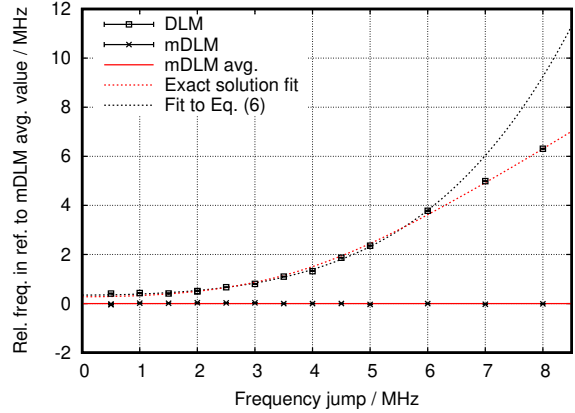}
    \caption{Dependence of the relative frequency at which the laser light was stabilized on $\Delta f$ in reference to mDLM average value (solid red line). Square points (DLM) are fit according to Eq.~(\ref{eq:zaleznosc_od_jrange}) (dashed black line) and to the solution of Eq.~(\ref{eq:rownanie_na_stabilizacje}) (dashed red line). The mDLM results (cross points) do not depend on the $\Delta f$.}
    \label{fig:jr}
\end{figure}

The background introduces a systematic error that cannot be avoided within DLM, even with  infinitesimally small frequency jumps $\Delta f$. The mDLM solves this issue as the spectral line with the slope background (Fig.~\ref{fig:bg})
\begin{figure}[ht]
    \centering
    \includegraphics[width=0.6\linewidth]{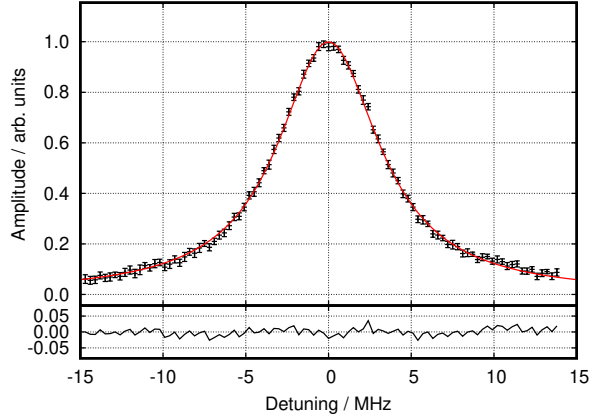}
    \caption{\textcolor{black}{Sub-Doppler spectral} line profile \textcolor{black}{obtained with the saturation spectroscopy} of the \printtransition transition in $^{201a}$Hg with linear background subtracted on-the-fly by the \textmu C. Subtracting the background makes the spectral line seen as symmetric. The results are averaged over fifteen scans and fit with the 
    Lorentzian
profile.
    The bottom part of the figure shows the residuals.}
    \label{fig:nobg}
\end{figure}
is seen to be symmetric due to on-the-fly background subtracting (Fig.~\ref{fig:nobg}). To validate the equation~(\ref{eq:zaleznosc_od_jrange}), the dependence of the frequency
at which the laser light was stabilized as a function of the frequency jump $\Delta f$ was measured for DLM and mDLM (Fig.~\ref{fig:jr}) with the \printtransition $^{201a}$Hg transition.

For the DLM, unlike the mDLM case, the frequency value shifts rapidly with the frequency jump. Moreover, even for a small $\Delta f$, the discrepancy \textcolor{black}{$\Delta_{\text{DLM-mDLM}}$} between results obtained with both methods persists. In general, this deviation depends on the ratio $a/A$ and the linewidth $w$. \textcolor{black}{Typically, $\Delta_{\text{DLM-mDLM}}$ is much smaller than the linewidth. Hence, for the Lorentzian profile, the deviation can be approximated as
\begin{equation}
    \Delta_{\text{DLM-mDLM}} \approx \frac{aw^2}{8A}.
\end{equation}
For our specific experimental conditions, the discrepancy shown in Fig.~(\ref{fig:jr}) is of $280$~kHz.}

\section{\textmu C software parameters}
In the mDLM, the frequency correction $f_{corr}$ is calculated by the \textmu C according to  equation~(\ref{eq:freq_corr}). The gain has to be adjusted accordingly to ensure the 
frequency correction
is fast enough to keep up with the frequency drifts of the spectral line and the laser light.
\begin{figure}[ht]
    \centering
    \includegraphics[width=0.6\linewidth]{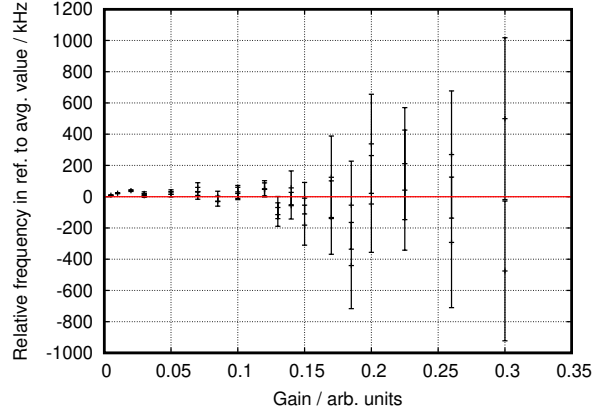}
    \caption{Dependence of the relative frequency at which the laser light was stabilized on the gain in reference to the mDLM average value (solid line). For high gain, the frequency corrections are too extensive,
    causing big uncertainties. Too low gains make the system unable to keep up with the frequency drifts of the laser light.}
    \label{fig:i}
\end{figure}
Fig.~\ref{fig:i} shows the dependence of the frequency at which the laser light was stabilized on the gain. For high gain, the frequency correction $f_{corr}$ is too extensive, which leads to high uncertainties. Too small gain makes the system unable to keep up with the frequency drifts, which results in scattering the results in the tens of kHz range. 

\begin{figure}[ht]
    \centering
    \includegraphics[width=0.6\linewidth]{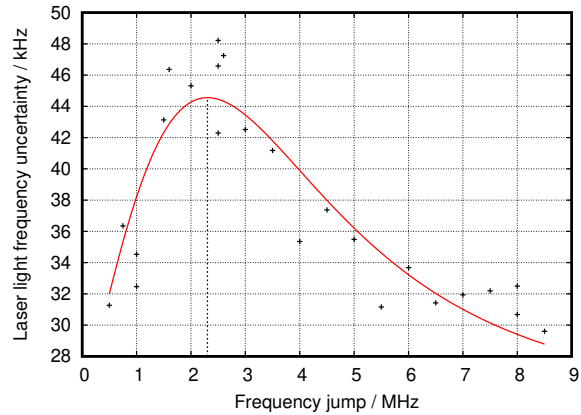}
    \caption{\textcolor{black}{Dependence of the laser light stabilizing frequency uncertainty on the frequency jump.}
The
first derivative of a Lorentz profile is fit to the collected data. 
    \textcolor{black}{The maximum of the fit corresponds to the $\Delta f = w/\sqrt{12}$ (dashed line), at which the mDLM sensitivity is the highest.}
}
    \label{fig:u_jrange}
\end{figure}

As shown in Fig.~\ref{fig:jr}, in the mDLM the frequency at which the laser light stabilizes
does not depend on the frequency jump $\Delta f$. 
Despite this,
the parameter $\Delta f$ should be adjusted according to the spectral linewidth \textcolor{black}{$w$}. 
For the optimal frequency jump, the sensitivity of the mDLM is the highest. This is the case for the steepest parts of the slopes being probed. It corresponds to the maximum of the first derivative of the spectral line profile. \textcolor{black}{For the Lorentzian profile, the optimal $\Delta f = w/\sqrt{12}$}. For this specific \textcolor{black}{value},
the deviation of the laser frequency from the spectral line's resonance results in the largest response of the correction system, shown in Fig.~\ref{fig:u_jrange}.

\section{Absolute frequency}
To validate the mDLM performance, we used it to determine the absolute frequency of the \printtransition transition in $^{201a}$Hg ($F = 3/2 \to F'=5/2$), located on the steepest slope of the Doppler profile among all Hg isotopes at room temperature.  
The determined absolute frequency is \textcolor{black}{$1181541111051(83)$~kHz}.
To obtain this value, the optical frequency comb was used
and 
the error budget was measured. 
\textcolor{black}{The total uncertainty was calculated according to the following formula}
\begin{equation}
    \textcolor{black}{u(f_{\text{abs}}) = \sqrt{u_{\text{stat}}^2 + u(\Delta_{\text{AC-Stark}})^2  + u(\Delta_{\text{Hg-Hg}})^2 + u(\Delta_{\text{Hg-H}_2})^2 + \Sigma^{3}_{i=1} u(\Delta_{\text{Zeeman}-i})^2},}
\end{equation}
\textcolor{black}{where $u_{\text{stat}}=3.9$~kHz is the statistical uncertainty of the line position measured under known and controllable experimental conditions, $\Delta_{\text{AC-Stark}}$ is the AC-Stark frequency shift, $\Delta_{\text{Hg-Hg}}$ and $\Delta_{\text{Hg-H}_2}$ are the pressure shifts, and $\Delta_{\text{Zeeman}-i}$ are Zeeman shifts of the $x$, $y$, and $z$-axis, respectively. All the components are described in the following subsections and the values are summarized in Table~\ref{tab:shifts}}.

\subsection{AC-Stark shift}
\begin{figure}[ht]
    \centering
    \includegraphics[width=0.6\linewidth]{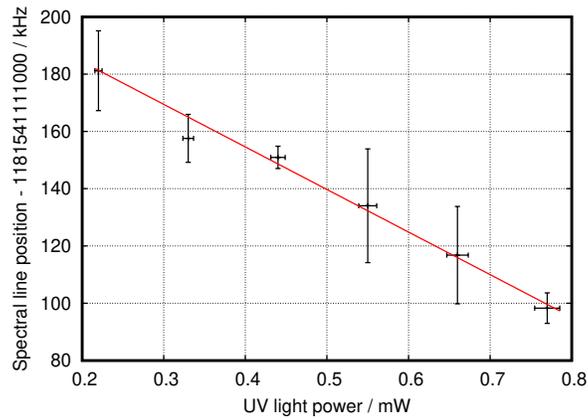}
    \caption{Dependence of the spectral line position on the total power of the UV beams. The 
solid red
line is a weighted fit to the results. The AC-Stark shift is deduced from the slope to be $-148.6(9.4)$~kHz/mW.}
    \label{fig:lightshift}
\end{figure}

The energy levels of the atoms are shifted under the influence of the electric field of the laser beam. To estimate the contribution of this effect, we measured the position of the spectral line for different UV light power (Fig.~\ref{fig:lightshift}). A linear fit weighted by the points' uncertainties yields the AC-Stark shift correction factor of $-148.6(9.4)$~kHz/mW. 

The uncertainties are mainly related to the accuracy of the power meter used (Thorlabs S130VC) and the power measuring method. The total uncertainty is deduced to be $5.4\%$.
\subsection{Pressure shift}
The
pressure-dependent collisions between Hg atoms and residual gases 
    affect the atomic transition frequency.
Since the vapor pressure of Hg depends on the temperature, the spectroscopic cell temperature was actively stabilized. The average temperature of the cell during measurements was \textcolor{black}{$19.97(50)~^\circ$C}, which corresponds to \textcolor{black}{$0.1708(75)$~Pa} vapor pressure~\cite{Hgpressure}. The temperature uncertainty is limited by the accuracy of the analog-digital converter used in the Pt100-based temperature measurement scheme.

We
used the same approach as presented in~\cite{Witkowski1} 
to estimate a collision-induced shift.
The collisions between mercury atoms shift the 
transition
frequency 
by $-22(22)$~kHz/Pa. 
The contribution from the collisions with residual gases was estimated, assuming molecular hydrogen as a dominant component, which is consistent with the fused silica cell residual gases composition~\cite{Palosz}. H$_2$ pressure at room temperature is less than $0.13$~Pa, 
corresponding
to the spectral line shift of $-35(35)$~kHz/Pa.

\subsection{Zeeman shift}
The $^{201}$Hg isotope (abundance of $13.18\%$ \cite{sklad}) is a fermion with non-zero nuclear spin ($I=3/2$), which leads to hyperfine splitting of the $^3P_1$ state into three states ($F = 5/2, 3/2, 1/2$)~\cite{HgMagnetometry}. The $F=3/2 \to F'=5/2$ transition is the measured one. 
The
ground state $^1S_0, F=3/2$ 
splitting is negligible 
in a weak magnetic field
as the nuclear Land\'e factor is much smaller than the orbital and the electron ones. The excited state $^3P_1, F=5/2$ splits into six substates characterized by magnetic quantum number $m_F$. To a first approximation, the linear Zeeman effect splits the sublevels symmetrically and therefore has no contribution to the absolute frequency shift. On the other hand, the quadratic Zeeman effect does not cause splitting but can produce a non-negligible shift of the hyperfine levels.

\begin{figure}[ht]
         \centering
         \includegraphics[width=0.6\linewidth]{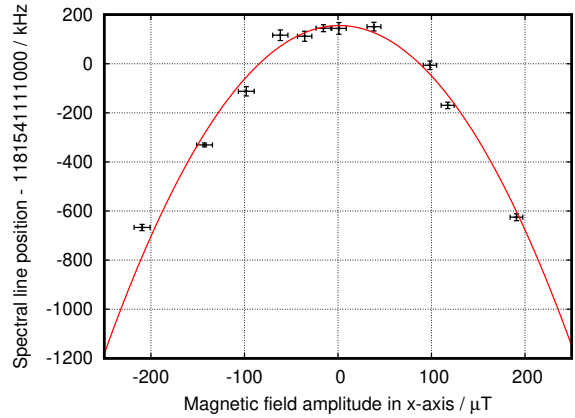}
         \caption{Dependence of the spectral line position  on the magnetic field in the $x$-axis (vertical, perpendicular to the UV beams). The solid red curve is a weighted parabola fit to the results. The quadratic and linear Zeeman shifts are deduced to be $-0.0211(13)$~kHz/\textmu T$^2$ and $0.07(16)$~kHz/\textmu T, respectively.}
         \label{fig:poleX}
\end{figure}

\begin{figure}[ht]
         \centering
         \includegraphics[width=0.6\linewidth]{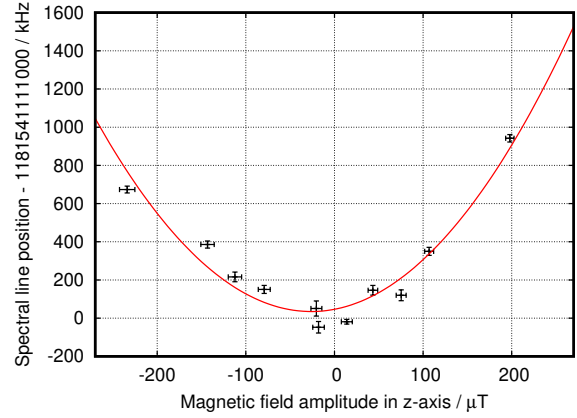}
         \caption{Dependence of the spectral line position  on the magnetic field in the $z$-axis (horizontal, perpendicular to the UV beams). The solid red curve is a weighted fit to the results. The quadratic and linear Zeeman shifts are deduced to be $0.0170(15)$~kHz/\textmu T$^2$ and $0.89(22)$~kHz/\textmu T, respectively.}
         \label{fig:poleZ}
\end{figure}

\begin{figure}[ht]
         \centering
         \includegraphics[width=0.6\linewidth]{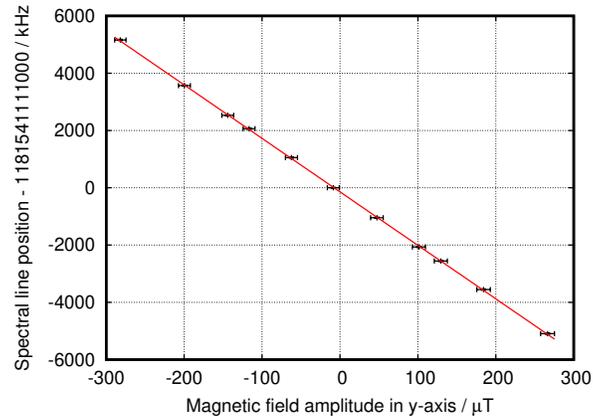}
         \caption{Dependence of the spectral line position on the magnetic field in the $y$-axis (along UV beams). The solid red line is a weighted fit to the results. The linear Zeeman shift is deduced to be \textcolor{black}{$-18.677(53)$~kHz/\textmu T}}
         \label{fig:poleY}
\end{figure}

To estimate the Zeeman shift, the spectral line position was independently  measured for different magnetic fields for each axis. The external magnetic field of the Helmholtz coils was varied over a range that did not split the spectral line measurably. The results corresponding to the vertical ($x$-axis, Fig.~\ref{fig:poleX}) and horizontal ($z$-axis, Fig.~\ref{fig:poleZ}),
both perpendicular to the UV beams, manifest significant quadratic  and linear Zeeman effects.
For the axial magnetic field component (Fig.~\ref{fig:poleY}), a very large shift, which exhibits only a linear dependence in the measured range, is observed.
The magnetic field was measured by eight 3-axis offset-compensated magnetometers located symmetrically around the spectroscopic cell. Each reading has an uncertainty related to the accuracy of the offset calibration. Each point in the graphs corresponds to the average value of the magnetometers' readings for a given axis. The variance of the readings specifies the uncertainties of the points.

To account for the Zeeman effect, the magnetic field and its gradient were measured during the absolute frequency measurement. The uncertainty-weighted sensors' readings were averaged,
resulting in six magnetic field values, one for each side of the spectroscopic cell. The uncertainty of these results was deduced from the calculated internal and external variances, where the larger of the two was chosen. The average of pairs of points along the same axis determines the value of the magnetic field at the center of the spectroscopic cell,
which was calculated to be $43.6(3.6)$~\textmu T, $-11.6(4.3)$~\textmu T,
and $-7.46(0.77)$~\textmu T for x, y, and z-axis, respectively.

\begin{table}[ht]
\centering
\caption{Systematic shifts and their uncertainties determined for the experimental conditions, i.e.,
UV beams total power of $0.440(24)$~mW, spectroscopic cell temperature of $19.97(50)~^\circ$C, and magnetic field of $43.6(3.6)$~\textmu T, $-11.6(4.3)$~\textmu T and $-7.46(0.77)$~\textmu T for x, y, and z-axis respectively. All results are in kHz.}
\label{tab:shifts}
\footnotesize
\begin{tabular}{cccccc} 
\hline
\multicolumn{6}{c}{\textbf{Effect}}                                                                                                                    \\ 
\hline
\textbf{AC-Stark} & \textbf{Pr. shift Hg-Hg} & \textbf{Pr. shift Hg-H$_2$} & \textbf{Zeeman x-axis} & \textbf{Zeeman y-axis} & \textbf{Zeeman z-axis}  \\ 
\hline
65.4(5.4)         & 3.8(3.8)                 & 4.6(4.6)                    & 37.1(9.8)              & -217(80)               & 5.7(1.7)                \\
\hline
\end{tabular}
\end{table}

\section{Conclusions}
In summary, we demonstrated a novel method to stabilize the frequency of 
laser light on the position of a spectral line.
The method is especially 
beneficial
in 
the
case of the atomic lines located on the slope of the absorption profile, which induces a systematic frequency shift. We derived a formula to calculate the background-induced shift, which was then compared with the experimental results, showing full agreement. To demonstrate the performance of our method, we performed a 
measurement of the absolute frequency of the \printtransition $^{201a}$Hg improving the accuracy by 
three orders of magnitude. Our result agrees with the previously measured value within the uncertainty.

\begin{backmatter}
\section*{Funding}
Narodowe Centrum Nauki (2021/42/E/ST2/00046, 2021/41/B/ST2/00681); European Metrology Programme for Innovation and Research (20FUN01 TSCAC).

\section*{Acknowledgments}
This project 20FUN01 TSCAC has received funding from the EMPIR programme co-financed by the Participating States and from the European Union’s Horizon 2020 research and innovation programme. 
This work was supported by the National Science Centre, Poland Project no. 2021/42/E/ST2/00046. A. Linek is supported by the National Science Centre, Poland, Project no. 2021/41/B/ST2/00681.

The measurements were performed as a part of research done at the National Laboratory FAMO (KL FAMO) in Toruń, Poland. KL FAMO in Toru\'n is supported by the subsidy of the Ministry of Science and Higher Education. The authors would like to thank Roman Ciury\l{}o for fruitful discussions. The authors also thank Mode-Locked Technology Sp. z o.o. for providing the digital phase detector for laser frequency stabilization.

\section*{Disclosures}
The authors declare no conflicts of interest.

\section*{Data availability}
Data underlying the results presented in this paper are available in the open data repository~\cite{repod}.

\end{backmatter}

\bibliography{pracki}

\end{document}